\documentclass[12pt]{article}
\usepackage{graphicx}
\usepackage{amssymb}
\usepackage{amscd}
\usepackage{amsmath}

\begin{document}
\begin{titlepage}

{\hbox to\hsize{\hfill January 2013 }}

\bigskip \vspace{3\baselineskip}

\begin{center}
{\bf \large 
Electroweak Vacuum (In)Stability in an Inflationary Universe}

\bigskip

\bigskip

{\bf Archil Kobakhidze and Alexander Spencer-Smith\\ }

\smallskip

{ \small \it
ARC Centre of Excellence for Particle Physics at the Terascale, \\
School of Physics, The University of Sydney, NSW 2006, Australia \\
E-mails: archilk@physics.usyd.edu.au, alexss@physics.usyd.edu.au
\\}

\bigskip
 
\bigskip

\bigskip

{\large \bf Abstract}

\end{center}
\noindent 
Recent analysis shows that if the 125-126 GeV LHC resonance turns out to be the Standard Model Higgs boson, the electroweak vacuum would be a metastable state at 98$\%$ C.L. In this paper we argue that, during inflation, the electroweak vacuum can actually be very short-lived, contrary to the conclusion that follows from the flat spacetime analysis. 
Namely, in the case of a pure Higgs potential the electroweak vacuum decays via the Hawking-Moss transition, which has no flat spacetime analogue. As a result, the Higgs vacuum is unstable, unless the rate of inflation is low enough: $H_{\rm inf}\lesssim 10^9-10^{12}$ GeV. Models of inflation with such a low rate typically predict negligible tensor perturbations in the cosmic microwave background radiation (CMBR). This is also true for models in which the perturbations are produced by a curvaton field.  We also find that if the effective curvature of the Higgs potential at a local maximum (which may be induced by inflaton-Higgs interactions) is large enough, then the decay of the electroweak vacuum is dominated by the Coleman-de Luccia transition. The electroweak vacuum is also short-lived in this case, due to a negative effective self-interaction coupling. Based on our analysis of Higgs vacuum stability during inflation, we conclude that the observation of tensor perturbations by the Planck satellite would provide strong indirect evidence for new physics beyond the Standard Model responsible for stabilisation of the electroweak vacuum.
 
 \end{titlepage}

\section{Introduction}

The discovery of a Higgs-like resonance at 125-126 GeV \cite{:2012gk}, \cite{:2012gu}, triggered much discussion on the stability of the electroweak vacuum. In the recent, most accurate analysis \cite{Degrassi:2012ry}, it was found that, if the Standard Model is valid up to Planckian  energies, a Higgs mass $m_h<126$ GeV implies that the electroweak vacuum is a metastable state at $98\%$ C.L. Any discussion of Higgs vacuum stability must be considered in a cosmological framework. The classical flat spacetime bounce solutions  \cite{Coleman:1977py},  \cite{Lee:1985uv}, which describe the transition from a false vacuum to a true vacuum in the semiclassical approximation, usually have small gravitational corrections \cite{Coleman:1980aw}, \cite{Isidori:2007vm}, and, in fact, the bound  quoted in \cite{Degrassi:2012ry} is considered in flat spacetime. 

There are strong theoretical arguments as well as indications from astrophysical observations in favour of an inflationary period of the early universe, which is well-approximated by a cosmological patch of de Sitter spacetime. Inflation not only solves some problems with the standard hot Big Bang theory, such as the horizon and flatness problems, but also provides a natural mechanism for the generation of large scale density (scalar) and gravitational wave (tensor) perturbations, imprinted as anisotropies in the cosmic microwave background radiation (CMBR)  \cite{Mukhanov:1990me}. False vacuum decay in (quasi-)de Sitter spacetime may differ dramatically from that in flat spacetime because there are known de Sitter bounce solutions which have no flat spacetime analogue and, moreover, in certain situations they dominate false vacuum decay processes. One of these solutions is the Hawking-Moss instanton \cite{Hawking:1981fz}. It usually provides the dominant contribution to false vacuum decay when a scalar potential around its local maximum, separating false and true vacua, is flat enough. In this paper we show that this is indeed the case for the Higgs potential and find that the electroweak vacuum is unstable during inflation, unless the inflationary rate, $H_{\rm inf}$, is low enough: $H_{\rm inf}\lesssim 10^9-10^{12}$ GeV. We also consider the situation in which large curvature of the Higgs potential is induced during inflation by inflaton-Higgs interactions. Alternatively, large effective curvature during inflation may be induced  due to the non-minimal coupling of the Higgs field to the gravitational scalar curvature.  In these cases we find that electroweak vacuum decay is dominated by the Coleman-de Luccia instanton, and the vacuum turns out to be unacceptably short-lived.

Interestingly, our analysis shows that important information about electroweak vacuum stability can be inferred from a possible observation of tensor fluctuations by the Planck satellite \cite{Planck:2006aa}. Such an observation would exclude low-scale inflationary models preferred by the requirement of vacuum stability, and would provide a strong indication of new physics responsible for stabilisation of the electroweak vacuum. \cite{Espinosa:2007qp} arrived at a similar conclusion in a previous publication, where electroweak vacuum stability during inflation was also discussed. However, the false vacuum decay mechanism considered there is due to the generation of large-amplitude quantum fluctuations of the Higgs field which, at super-horizon scales, behave as a homogeneous classical field. This mechanism does not work when effective curvature of the Higgs potential is larger than the inflationary rate. Also, the arguments given in \cite{Espinosa:2007qp} are not straightforwardly applicable when the density perturbations are generated by fields other than the inflaton. Our work is based on a quantitatively different physical picture and  therefore, we believe that the study presented in this paper provides an important extension of \cite{Espinosa:2007qp}.   

The rest of the paper is organised as follows. In the next section we briefly review electroweak vacuum stability in flat spacetime. In section 2 we discuss Hawking-Moss and Coleman-de Luccia transitions in de Sitter spacetime, followed by a section in which we discuss constraints on models of inflation from the requirement of electroweak vacuum stability.
         
\section{Electroweak vacuum stability in flat spacetime}

The decay rate of the electroweak vacuum is calculated as follows.  Since the instability scale (which is defined as the scale $\mu_{\rm i}$ where the running  Higgs quartic coupling $\lambda(\mu)$ changes sign from positive to negative, that is, $\lambda(\mu_{\rm i})=0$) is much higher than the electroweak scale: $\mu_{\rm i}\gg v_{\rm EW}$, one can neglect the Higgs mass parameter and thus approximate the 1-loop Higgs potential as \cite{Altarelli:1994rb}:
\begin{equation}
V_{\rm H}=\frac{\lambda_{\rm eff}(h)}{4}h^4~,
\label{1}
\end{equation}    
where $\lambda_{\rm eff}(h)=\lambda(\mu)+\beta_{\lambda}\ln\left(\frac{h}{\mu}\right)$. Here we ignore the numerically less significant quantum correction arising from wave-function renormalisation and $\beta_{\lambda}$ is the 1-loop $\beta$-function that defines the running of the Higgs quartic coupling, $\lambda(\mu)$. Since the potential \eqref{1} does not depend on the renormalisation scale, $\mu$, we find it convenient to evaluate the potential at the instability scale, $\mu_i$. Hence we have:
\begin{equation}
V_{\rm H}=\frac{\bar\beta_{\lambda}\ln\left(\frac{h}{\mu_{\rm i}}\right)}{4}h^4~,
\label{2}
\end{equation}    
where $\bar \beta_{\lambda}$ is the $\beta_{\lambda}$ at $\mu=\mu_i$. 

\begin{figure}
\centering
\includegraphics[width=0.8\textwidth]{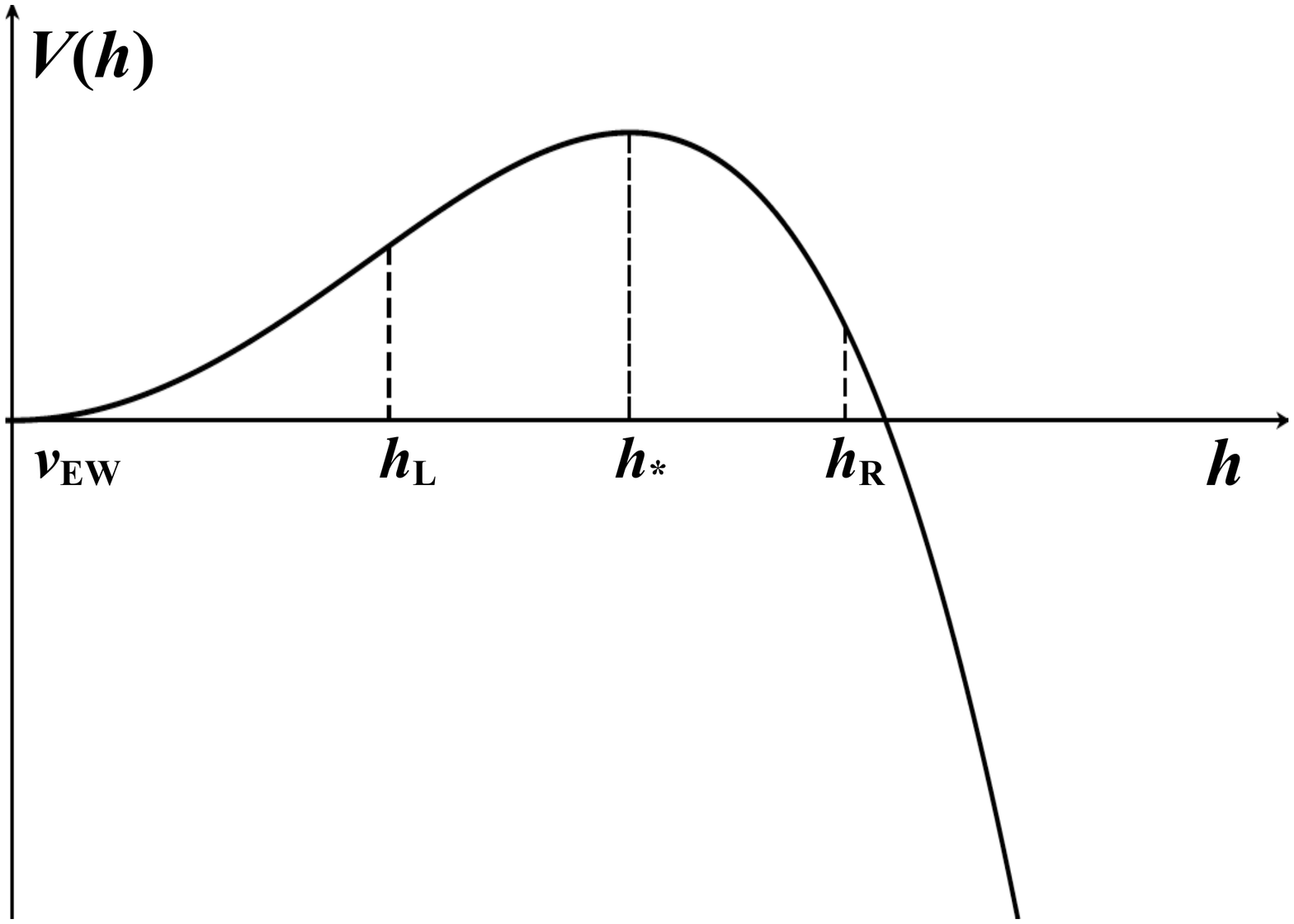}
\vspace{-1cm}
  \caption{\small The Higgs potential. For large values of the Higgs field h, the electroweak vacuum configuration is regarded as trivial,  $v_{EW}\approx 0$. }
  \label{fig}
\end{figure}

The stationary equation, $\frac{dV_{\rm H}}{dh}=0$, admits two solutions. One is the local minimum, $h_{\rm fv}=0$, which, in the given approximation ($v_{\rm EW}\approx 0$), represents the electroweak vacuum configuration. Another, $h_{*}=\mu_{\rm i} {\rm e}^{-1/4}$, represents a local maximum since $\bar \beta_{\lambda}<0$ in the Standard Model and thus the curvature of the potential, $\mu_{*}$, at $h=h_{*}$, is negative: $\mu_{*}^2\equiv\left.\frac{d^2V_{\rm H}}{dh^2}\right\vert_{h=h_{*}}=\frac{\bar \beta_{\lambda}}{16\sqrt{e}}\mu_{\rm i}^2<0$.  The potential is sketched in Fig. \ref{fig}. 

In flat spacetime the transition from the electroweak false vacuum to a deeper (negative energy) true vacuum is described by the $O(4)$-symmetric Lee-Weinberg bounce \cite{Lee:1985uv}. The classical Euclidean action evaluated at this configuration is:
\begin{equation}
S_{\rm LW}=\frac{8\pi^2}{3\vert \lambda(\mu_m)\vert}~,
\label{3}
\end{equation}   
 where the renormalisation scale, $\mu_m$, ($R\sim 1/\mu_m$ is the size of the bounce) is that which minimises the action \eqref{3}, i.e. $\beta_{\lambda}(\mu_m)=0$ \cite{Isidori:2001bm}. In the Standard Model $\mu_m\approx 10^{17}$ GeV. Hence, very small bounces ($R\mu_i\ll1$) provide the dominant contribution to false vacuum decay and, therefore, spacetime curvature effects are expected to be small (this was demonstrated explicitly in \cite{Isidori:2007vm}). Thus the transition probability within the lifetime of the universe, $\tau_{\rm U}=1/H_{0}\approx 10^{32}$ eV$^{-1}$, is\footnote{Actually, ${\rm e}^{-p}$ gives the probability to find the visible universe in the false vacuum, while $(1-{\rm e}^{-p})$ is the probability of having the universe in the true vacuum.}:
\begin{equation}
p=\left(\frac{\mu_m}{H_0}\right)^4 
{\rm e}^{-S_{\rm LW}}\approx {\rm e}^{390-\frac{26.3}{\vert \lambda(\mu_m)\vert}}\ll1~,
\label{4}
\end{equation} 
since $\vert \lambda(\mu_m)\vert \approx 0.01-0.02$ depending on the top quark mass\footnote{The Higgs vacuum stability is very sensitive to the top quark mass. Here we take $m_t=173\pm 0.9$ GeV \cite{Lancaster:2011wr}. However, as it has been pointed out in \cite{Alekhin:2012py}, uncertainties in top quark mass measurements may actually be larger.}. Therefore the electroweak vacuum can be regarded as a metastable, but safely long-lived state  for $m_h\approx 125-126$ GeV \cite{Degrassi:2012ry}.

\section{Electroweak vacuum stability during inflation}
 
Vacuum stability must obviously be discussed in a cosmological context, where the question of initial conditions becomes important.  \emph{A priori}, the magnitude of the Higgs field can assume any value in the interval $[0, M_{\rm P}]$, so the probability of finding it at an initial value less than the instability scale, $h\lesssim \mu_{\rm i}\sim 10^{10}$ GeV, is, in fact, extremely low: $~(\mu_{\rm i}/M_{\rm P})^4\approx 10^{-32}$ \cite{Hertzberg:2012zc}.  We recognise this as a serious problem, but take, nevertheless, a tolerant attitude. I.e. we assume that, as a result of some unknown dynamical mechanism (see, e.g. \cite{Lebedev:2012sy}) or via anthropic reasoning, the Higgs field is initially in its false electroweak vacuum state.   

Inflation is a widely accepted paradigm favoured by some astrophysical observations. Accepting the inflationary paradigm, we extend our Higgs potential \eqref{1} by adding terms containing the inflaton field $\phi$:
\begin{equation}
V(h,\phi)=V_{\rm inf}(\phi)+\bar V(\phi, h)~,~{\rm where~}\bar V(\phi, h)=V_{\rm H}(h)+V_{\rm H-inf}(h,\phi)
\label{5}
\end{equation}   

It is convenient to expand the inflaton potential in a Taylor series around $\phi_{\rm inf}$, the field value just at the start of the observable inflation era, that is, $\sim N=60$ e-folds before the  end of inflation when the density perturbation on the scale of present Hubble radius was generated:
\begin{equation}
V_{\rm inf}(\phi)={\cal V}_{\rm inf}+V'_{*}(\phi-\phi_{\rm inf})+\frac{1}{2}V''_{*}(\phi-\phi_{\rm inf})^2+...,
\label{6}
\end{equation} 
where ${\cal V}_{\rm inf}\equiv V(\phi_{\rm inf})$, $V'_{*}\equiv \left. dV/d\phi \right\vert_{\phi=\phi_{\rm inf}}$, $V''_{*}\equiv \left. d^2V/d\phi^2 \right\vert_{\phi=\phi_{\rm inf}}$, etc. We assume that $\bar V(h=0,\phi)=0$ without loss of generality. The potential \eqref{5} must be flat enough around $(\phi=\phi_{\rm inf}, h=0)$, i.e.,  
\begin{equation}
\epsilon \equiv \frac{M_{\rm P}^2}{2}\left(\frac{V'_{*}}{{\cal V}_{\rm inf}}\right)^2\ll 1~,~~-1\ll\eta \equiv M_P^2\frac{V''_{*}}{{\cal V}_{\rm inf}}\ll 1~,
\label{7}
\end{equation}
so that the energy density in the universe is dominated by the constant term: $V_{\rm inf}\approx {\cal V}_{\rm inf}$. This leads to a de Sitter-like exponential expansion of the universe characterised by the expansion rate $H_{\rm inf}=\sqrt{\frac{{\cal V}_{\rm inf}}{3M^2_{\rm P}}}$, where $M_{\rm P}\approx 2.44\cdot 10^{18}$ GeV is the reduced Planck mass.  Models of inflation are broadly classified as large-field ( $2\epsilon>\eta>0$), small-field ($\epsilon>0,~\eta<0$) and hybrid ($\eta>2\epsilon$).

To discuss instanton mediated transitions between false and true vacua we consider the Euclidean analogue of cosmological spacetime:
\begin{equation}
ds^2=d\chi^2+\rho^2(\chi)d\Omega_3^2~,
\label{8}
\end{equation}
where $d\Omega_3$ is the length element on a 3-sphere, $\rho(\chi)$ is the Euclidean scale factor and $\chi^2=t^2+r^2$. The relevant Euclidean action for homogeneous fields is:
\begin{eqnarray}     
S=\int\sqrt{g}d^4x \left[-\frac{M_{\rm P}^2}{2}{\cal R}+\frac{1}{2}\partial_{\mu}\phi\partial^{\mu}\phi+
\frac{1}{2}\partial_{\mu}h\partial^{\mu}h+V\right] \nonumber \\
=2\pi^2\int d\chi \rho^3\left[
\frac{1}{2}\dot{\phi}^2+\frac{1}{2}\dot{h}^2+V+\frac{3M_{\rm P}}{\rho^2}\left(\dot{\rho}^2+\rho\ddot\rho -1\right)
\right]~,
\label{9}
\end{eqnarray}
where $\dot{A}\equiv dA/d\chi$ and $\ddot{A}\equiv d^2A/d\chi^2$.  The resulting equations of motion read:
\begin{eqnarray}
\label{10}
\ddot{\phi}+3\frac{\dot{\rho}}{\rho}\dot{\phi}=\frac{\partial V}{\partial \phi}~, \\
\label{11}
\ddot{h}+3\frac{\dot{\rho}}{\rho}\dot{h}=\frac{\partial V}{\partial h}~, \\
\dot{\rho}^2=1+\frac{1}{3M_{\rm P}^2}\rho^2\left(\frac{1}{2}\dot{\phi}^2+\frac{1}{2}\dot{h}^2 -V\right)~. 
\label{12}
\end{eqnarray}
We take $\phi\approx \phi_{\rm inf}=const.$ as an approximate solution to Eq.~\eqref{10} during inflation. Assuming that the inflaton provides the dominant contribution to the energy of the system, $V\approx {\cal V}_{\rm inf} \gg \dot h^2$, it is easy to verify that Eq.~\eqref{12} admits the solution $\rho=H_{\rm inf}^{-1}\sin\left(H_{\rm inf}\chi\right)$ ($\chi \in [0,\pi/H_{\rm inf}]$).  Hence, in this fixed background approximation we are left with the following equation for $h$:
\begin{equation}
\ddot h+3H_{\rm inf}\cot\left(H_{\rm inf}\chi\right)\dot h=\frac{\partial V(h,\phi_{\rm inf})}{\partial h}~. 
\label{13}
\end{equation}
Since $\cot\left(H_{\rm inf}\chi\right)$ diverges at $ \chi= \{0, \pi/H_{\rm inf} \}$, we supplement equation \eqref{13} with the boundary conditions: $\dot h(0)=\dot h(\pi/H_{\rm inf})=0$.

\subsection{Vacuum decay via the Hawking-Moss instanton}

Eq.~\eqref{13} admits the trivial solution, $h=0$, which describes the electroweak false vacuum state during inflation. Noting that $\left.\frac{\partial V(\phi_{\rm inf}, h)}{\partial h}\right \vert_{h=h_{*}}=0$, we find another simple solution:  $h=h_{*}$. The transition between the electroweak vacuum state $(\phi_{\rm inf}, h=0)$ and the state $(\phi_{\rm inf}, h_{*})$ corresponding to the top of the barrier in the $h$-direction (see Fig. \ref{fig}) during inflation is determined by the difference between the corresponding Euclidean actions:
\begin{equation}
B_{\rm HM}=S[\phi=\phi_{\rm inf}, h=h_{*}]-S[\phi=\phi_{\rm inf}, h=0]=\frac{8\pi^2}{3H_{\rm inf}^4}\left(V(\phi_{\rm inf},h_{*})-{\cal V}_{\rm inf} \right)~.
\label{14}
\end{equation}  
This is the Hawking-Moss transition, which typically dominates in false vacuum decay processes when the curvature of the potential near it's local maximum is smaller than the expansion rate during inflation \cite{Hawking:1981fz}. Physically, the Hawking-Moss transition can be understood as a thermally-driven transition where the false vacuum state, $h=0$, is excited to unstable state, $h=h_{*}$ within the Hubble radius \cite{Brown:2007sd}. The de Sitter spacetime plays the role of a thermal reservoir with associated Hawking-Gibbons temperature $T_{\rm dS}=H_{\rm inf}/2\pi$ \cite{Brown:2007sd}. Once $h=h_*$ there is unit probability that the configuration will subsequently roll down towards the negative energy true vacuum state. Since the presently observable universe was within the Hubble horizon at the start of inflation, $r_0<1/H_{\rm inf}$, the fast decay of the electroweak vacuum into a negative energy ($\sim M_{\rm P}^4$) true vacuum within the entire Hubble patch   prevents inflation from proceeding and, in fact, results in a disastrous collapse of the universe into a black hole singularity (`Big Crunch') rather than relaxation into the true vacuum. The transition probability from the electroweak vacuum $(\phi_{\rm inf}, 0)$ to a state $(\phi_{\rm inf}, h_{*})$ at the start of inflation (during one Hubble time $\tau =1/H_{\rm inf}$), 
 is given by:
\begin{equation}
p\approx (r_0H_{\rm inf})^{-4}{\rm e}^{-B_{\rm HM}}\approx \exp\left\lbrace -
\frac{8\pi^2}{3}\frac{V_{\rm H}(h_{*})+V_{\rm inf-H}(\phi_{\rm inf},h_{*})}{ H^4_{\rm inf} }
\right\rbrace~,
\label{15}
\end{equation}   
where we conservatively assumed that the size of the presently observable universe was $r_0\sim 1/H_{\rm inf}$ at the start of inflation. 

\subsection{Vacuum decay via the Coleman-de Luccia instanton}     

Besides the unknown inflation rate, $H_{\rm inf}$, the above probability also depends on unknown inflaton-Higgs interactions described by $V_{\rm inf-H}$. It is reasonable to require that these interactions do not spoil the flatness of the inflaton potential, $V_{\rm inf}$, in order to satisfy the slow-roll conditions \eqref{7}, as well as the flatness of the Higgs potential, $V_{\rm H}$, in order to keep the Higgs mass light. However, if we introduce a large bare mass for the Higgs boson and accept an unnatural fine-tuning of parameters, the effective curvature of the Higgs potential near the local maximum can be large as a result of inflaton-Higgs interactions. In that case the Hawking-Moss transition is suppressed and false vacuum decay via the Coleman-de Luccia instanton becomes dominant. To make our discussion quantitative let us consider the following inflaton-Higgs interaction potential: $V_{\rm inf-H}=\frac{1}{2}f(\phi)h^2$, where $f(\phi)$ is an arbitrary positive function of the inflaton field $\phi$. In the fixed background approximation ($\phi = const.$) this interaction potential turns into an effective mass term for the Higgs field, with $m_{\rm eff}^2\equiv f(\phi_{\rm inf})$. Then the total effective Higgs potential during inflation can be written as:
\begin{equation}
\bar V(\phi_{\rm inf}, h)=\frac{1}{2}m_{\rm eff}^2h^2\left[1 -\frac{1}{2}\left(\frac{h}{h_{*}}\right)^2\right]~,
\label{16}
\end{equation}    
where $h_{*}$ is a local maximum,
\begin{equation}
h_{*}=\sqrt{-\frac{m_{\rm eff}^2}{\lambda}}~,
\label{17}
\end{equation}
for $m_{\rm eff}^2>0$, and recall that for large values of $h$ we have $\lambda<0$ (the potential is unbounded from below, see Fig. {\ref{fig}}) so here we approximate it as a constant.
 
We are interested in the case of large effective curvature of the potential at the local maximum, that is, $m^2_{\rm eff}\gg\frac{1}{2}H_{\rm inf}^2$. In this limit Eq.~\eqref{13} takes the form of the corresponding flat spacetime equation:
\begin{equation}
h''+\frac{3}{x}h'=\frac{\partial \bar V(\phi_{\rm inf},h)}{\partial h }~,
\label{18}
\end{equation}
where $A'\equiv dA/dx$, $A''\equiv d^2A/dx^2$ and $x=m_{\rm eff}\chi$. Decay of the electroweak false vacuum proceeds through the nucleation of a bubble, with the field $h(x)$ approaching $h_{\rm R}>h_{*}$ near the centre and $h_{\rm L}<h_{*}$ far outside the bubble (see Fig. \ref{fig}). Note that, contrary to the similar process of the bubble nucleation in flat spacetime, $h_{\rm L}$/$h_{\rm R}$ need not necessarily be the false/true vacuum in de Sitter spacetime \cite{Brown:2007sd}.  For the interior bubble solution we approximate the potential \eqref{16} by neglecting the quartic coupling, while for the exterior solution  we neglect the quadratic coupling instead. These two solutions must be continuous at $x_{*}$:
\begin{equation}
h(x_*)=h_*~.
\label{19}
\end{equation}           

The following one-parameter class of bounce solutions can be obtained analytically in the above approximation:
\begin{eqnarray}
\label{20}
h(x)=\left\lbrace 
\begin{tabular}{cc}
$8h_{\rm R}\left(8+\left(\frac{h_{\rm R}}{h_{*}}\right)^2x^2\right)^{-1}~,$ & $0\leq x < x_{*}$ \\ 
$\frac{x_*h_*}{x\left(J_1(ix_*)+iY_1(-ix_*)\right)}\left(J_1(ix)+iY_1(-ix)\right)~,$& $x_{*}<x<\infty$ \\ 
\end{tabular} 
\right. ~, \\
x_*=\frac{2\sqrt{2}h_*}{h_{\rm R}}\left(\frac{h_{\rm R}}{h_*}-1\right)^{1/2}~.
\label{21}
\end{eqnarray}
Then we can compute the exponential B-factor:
\begin{equation}
B_{\rm CdL}=-\frac{2\pi^2}{\lambda}I~,
\label{22}
\end{equation}
where
\begin{equation}
I=\int_0^{\infty}x^3 dx\left[h^2(x)\left(1-\frac{h^2(x)}{2h_*^2}
\right)\right]~.
\label{23}
\end{equation} 


It should be stressed that fast false vacuum decay ($p\gtrsim 1$ per unit Hubble time ) quickly ceases inflation not only within a Hubble volume but also globally. Indeed, the fraction of an initial Hubble-volume that is still inflating after time $\tau$ is ${\rm e^{3H_{\rm inf}\tau}}{\rm e}^{-(\tau H_{\rm inf})^4p}$. The time at which inflation stops globally is then $\tau_{\rm stop}=\left(3/p\right)^{1/3}/H_{\rm inf}$, that is, $\approx 1.4$-Hubble time or less. Therefore, it is justified to derive constraints on inflation from vacuum instability by considering a single Hubble patch only.    

\section{Constraints on inflation from vacuum stability}

First we consider the case of negligible inflaton-Higgs interactions, i.e we set $V_{\rm inf-H}=0$ in \eqref{5}. In this case $V_{\rm H}(h_{*})$ is given by \eqref{2}, where $h_{*}=\mu_{\rm i}{\rm e}^{-1/4}$, and decay of the electroweak vacuum proceeds via the Hawking-Moss instanton, with $p$ given by \eqref{15}. The condition of vacuum (meta)stability, $e^{-p}>1/2$, implies the following constraint on the rate of inflationary expansion:
\begin{equation}
H_{\rm inf}<1.7\cdot 10^{9} (1.0\cdot 10^{12})~ {\rm GeV}~,
\label{24}
\end{equation}
for $m_h=126$ GeV and $m_t=174 (172)$ GeV. For $m_h=125$ GeV the limits are an order of magnitude lower.  Assuming that quantum fluctuations of the inflaton field are entirely responsible for the production of density perturbations that are imprinted in the CMBR temperature anisotropy, we can compute the standard relations between various observables and the slow-roll parameters \eqref{8}. In the linear approximation they are:
\begin{eqnarray}
\label{25}
{\cal A}_{s}=(8\pi^2\epsilon)^{-1/2}\frac{H_{\rm \inf}}{M_{\rm P}}~, \\
\label{26}
n_s=1-6\epsilon +2\eta ~, \\
\label{27}
r=16\epsilon~,
\end{eqnarray} 
where ${\cal A}_{s}$ is the amplitude of scalar perturbations, $n_s$ is the spectral index and $r$ is the ratio of squares of scalar and tensor perturbation amplitudes. Current experimental values for the above observables are \cite{Story:2012wx}:
\begin{eqnarray}
\label{28}
{\cal A}_{s}^{\rm exp}\approx 5\cdot 10^{-5}~, \\
\label{29}
n_s^{\rm exp}=0.9538 \pm 0.0081~, \\
\label{30}
r^{\rm exp}< 0.11~,
\end{eqnarray} 

As follows from \eqref{25}, in order to reproduce the observed amplitude of density perturbations \eqref{28} with the inflation scale \eqref{24}, one must have $\epsilon \approx 10^{-12} (10^{-6})$. The observed negative tilt ($n_s<1$) in the spectral index \eqref{29} then implies $\eta<0$ and, therefore, only small-field models are viable. Note that the tensor-to-scalar ratio is also tiny in this case: $r\approx 10^{-11}(10^{-5})$, well below the sensitivity ($\sim 10^{-2}$) of the Planck satellite experiment \cite{Planck:2006aa} to CMBR B-modes. Therefore an observation of tensor perturbations by Planck would rule out small-field inflationary inflationary models, leaving the problem of vacuum stability unresolved within the Standard Model.

As discussed in the previous section, sizeable inflaton-Higgs interactions, $V_{\rm inf-H}$, may increase the effective curvature of the scalar potential near the local maximum, $h=h_{*}$, during inflation and thus reduce the Hawking-Moss transition probability \eqref{15}. This possibility was recently discussed in \cite{Lebedev:2012sy} as a mechanism to ensure vacuum stability during inflation. Following \cite{Lebedev:2012sy} consider, for example, $V_{\rm inf-H}=\frac{\alpha}{2} h^{2}\phi^{2}$ (with $\alpha >0$). Then the effective mass, $m_{\rm eff}$, during inflation, is given by $m^2_{\rm eff}=\alpha\phi_{\rm inf}^2$ and $h_{*}=\left(-\alpha /\lambda\right)^{1/2}\phi_{\rm inf}>\mu_{\rm i}{\rm e}^{-1/4}$ ($\lambda(h_{*})<0$), so that the transition probability \eqref{15} is suppressed by the large negative factor in the exponent. This can be achieved in large-field ($\phi_{\rm inf}> M_{P}$) chaotic inflation models providing
 \begin{equation}
\alpha > 1.4\sqrt{|\lambda|}\left(\frac{H_{\rm inf}}{\phi_{\rm inf}}\right)^2~.
\label{31}
\end{equation}
It is interesting that current data \cite{Komatsu:2010fb}, \cite{Story:2012wx}  disfavors  all large-field models except the simplest one with quadratic potential\footnote{The model with a quadratic potential can also be viewed as an approximation of the model of natural inflation \cite{Freese:1990rb}.}: $V_{\rm inf}=\frac{1}{2}m_{\phi}^2\phi^2$, where the inflaton mass is $m_{\phi}\approx 10^{-5}M_{\rm P}$. Taking $|\lambda|\approx 0.01$, we obtain $\alpha > 6.0\cdot 10^{-12}$ within this model. It was estimated in \cite{Lebedev:2012sy} that, to have a sufficiently flat quantum-corrected inflaton potential, one needs to assume $\alpha <10^{-3}$. However, quantum corrections from inflaton-Higgs interactions also give a contribution to the Higgs potential, i.e. the Higgs mass receives a finite radiative correction, which in one-loop approximation reads:
 \begin{equation}
\delta m_h^2\approx \frac{\alpha}{64\pi^2}m_{\phi}^2~.
 \label{32}
 \end{equation}       
By demanding that this correction not exceed the observed Higgs mass: $\delta m_h<m_h$, we obtain $\alpha<64\pi^2 (m_h/m_{\phi})^2\approx 1.7\cdot 10^{-20}$, which is incompatible with the above vacuum (meta)stability condition \eqref{31}. Note that a very small $\alpha$ parameter is technically natural, because in the limit $\alpha \to 0$ the Higgs and inflaton sectors decouple (see, for example, the discussion in \cite{Foot:2007iy}).

However, if we abandon the naturalness condition, we can always introduce a bare Higgs mass parameter and tune it with radiative corrections to obtain the observed Higgs mass without further restriction on the strength of the inflaton-Higgs interaction. In this case the contribution to false vacuum decay is dominated by the Coleman-de Luccia transition described by Eq.~\eqref{22}. Numerically, we find that the integral \eqref{23} is dominated by a negative contribution from the interior solution,  i.e., $I<0$. Since the self-interaction coupling is also negative, the bubble nucleation rate is $p\sim {\rm e}^{-B_{\rm CdL}}\gg 1$, and thus the electroweak vacuum is again unstable.       

Finally, there is also the possibility that CMBR perturbations are produced by quantum fluctuations of a field other than the inflaton field, as in the curvaton scenario \cite{Enqvist:2001zp}. Then the observables \eqref{25}-\eqref{27} are modified to \cite{Moroi:2005np}:
\begin{eqnarray}
\label{33}
{\cal A}'_{s}= \sqrt{{\cal A}_{s}^2+\frac{g^2(X)H^2_{\rm inf}}{8\pi^2M^2_{\rm P}}}~,\\
\label{34}
n'_s=1-2\epsilon -\frac{4\epsilon-2\eta}{1+\epsilon g^2(X)}~,\\
\label{35}
r'=\frac{16\epsilon}{1+\epsilon g^{2}(X)}~,
\end{eqnarray} 
where $g^2(X)=\frac{3X}{2}$ for a `large' initial curvaton field value, $\sigma_{*}$, with $X=\sigma_{*}/M_{\rm P}\gg1$, and $g^2(X)=\frac{2}{X}$ for a `small' initial curvaton field: $X\ll1$. With the above equations we see that, if the curvaton is a major source of density perturbations then $g\approx 10^3-10^6$ is required for low scale inflation models. This again implies vanishing tensor perturbations: $r'\approx 0$. Thus we come to the conclusion based on the above vacuum stability analysis: an observation of tensor perturbations using the upcoming data collected by the Planck satellite would provide a strong hint of new physics responsible for stabilisation of the electroweak vacuum. This conclusion applies to all types of popular inflationary models discussed in the literature and even the cases where the effective curvature of the Higgs potential is large ($m_{\rm eff}>H_{\rm inf}$).

\section{Conclusion}

We have considered the question of electroweak vacuum stability in an inflationary universe.  We found that electroweak vacuum decay is governed by the Hawking-Moss transition and it is unacceptably fast, unless the rate of inflation is  $H_{\rm inf}\lesssim 10^9-10^{12}$ GeV. We also considered inflaton-Higgs interactions which may induce large curvature in the effective Higgs potential during inflation, in which case vacuum decay is dominated by the Coleman-de Luccia transition. We have demonstrated that the electroweak vacuum is also unstable in this case, essentially due to the negative effective Higgs self-interaction coupling.    

These observations have an interesting implication for CMBR fluctuations. Models of inflation with a low inflationary rate typically predict very small amplitude tensor fluctuations, with the same conclusion following from curvaton models. Thus, the upcoming data from the Planck satellite may provide important information concerning stability of the electroweak vacuum. Namely, the observation of tensor fluctuations would provide strong indirect evidence in favour of new physics beyond the Standard Model, responsible for stabilisation of the electroweak vacuum.

\paragraph{Acknowledgments.} This work was partially supported by the Australian Research Council.

\end{document}